\documentclass{webofc}
\usepackage[varg]{txfonts}   
\usepackage{bm}
\begin{document}
\title{Dynamics of causal hydrodynamic fluctuations \\ in an expanding system}
%
%

\author{\firstname{Shin-ei} \lastname{Fujii}\inst{1}\fnsep\thanks{\email{s-fujii-8c2@eagle.sophia.ac.jp}}
\and
\firstname{Tetsufumi} \lastname{Hirano}\inst{1}\fnsep\thanks{\email{hirano@sophia.ac.jp}}
}

\institute{Department of Physics, Sophia University, Tokyo 102-8554, Japan}

\abstract{%
We develop a framework of causal hydrodynamic fluctuations in one-dimensional expanding system  performing linearisation of the hydrodynamic equations around the boost invariant solution.
Through the description of space-time evolution of thermodynamic variables and flow velocity, we find a novel phenomenon that the structure of thermodynamic variables is almost frozen.
We also show that two-particle correlation functions of final hadrons after freezeout are closely related with the mass of hadrons and properties of the medium such as viscosity, relaxation time and equation of state.
}

\maketitle

\section{Introduction}
\label{intro}
Nowadays, relativistic dissipative hydrodynamics including shear and/or bulk viscosities has been used to extract transport properties of the QGP from experimental data.
According to the fluctuation-dissipation relations (FDR), hydrodynamic fluctuations and dissipations are, however, always accompanied with each other.
Since phenomena induced by the hydrodynamic fluctuations include the information of transport coefficients through FDRs, these enable us to analyse transport properties of the QGP from a viewpoint of hydrodynamic fluctuations.
Therefore, FDRs require hydrodynamic fluctuations in the dynamical framework of relativistic heavy ion collisions.

The first application of hydrodynamic fluctuations to the phenomenology of relativistic heavy ion collisions was made in Ref.~\cite{Kapusta} and later extended to causal framework in Ref.~\cite{Chattopadhyay}.
They linearised the hydrodynamic equations around the boost invariant solution in one-dimensional expanding system \cite{Bjorken} and regarded the linearised equations  as equations of motion (EoM) for fluctuations. 
Solving these equations, they obtained correlation of pion yield fluctuations as a function of rapidity gap.
Although the basic idea of our work is almost the same as Ref.~\cite{Chattopadhyay}, we focus more on analysis of properties of the QGP, identified hadron spectra and event-by-event phenomena induced by hydrodynamic fluctuations.

Although our formalism is almost the same as the one obtained in Ref.~\cite{Chattopadhyay}, we regard the linearised equations as stochastic differential equations and solve them numerically on an event-by-event basis.
We demonstrate the event-by-event space-time evolution of fluctuations of thermodynamic variables and analyse the two-point correlation functions of them.
To see the effects of hydrodynamic fluctuations on experimental observables, we also analyse two-particle correlation functions of final hadrons.

\section{Model}
\label{sec-1}
To perform linearisation, we first assume small deviations of four flow velocity, $u^\mu$, from the boost invariant solution \cite{Bjorken},
\begin{equation}
u^{\mu} = \left( \cosh \left( \eta_{\mathrm{s
}} + \delta y(\tau,\eta_{\mathrm{s}}) \right), 0, 0, \sinh \left(\eta_{\mathrm{s}} + \delta y(\tau,\eta_{\mathrm{s}})\right) \right), \label{eq:velocity}
\end{equation}
where $\eta_{\mathrm{s}} \equiv \tanh^{-1} (z/t)$ and $\delta y$ are space-time rapidity and deviation of flow rapidity from boost invariant solution, respectively.
Correspondingly, all thermodynamic variables in the energy-momentum tensor can be expanded as, e.g., $e \approx e_{0}(\tau) + \delta e(\tau,\eta_{\mathrm{s}})$.
All variables with subscripts $0$ denote variables of background and these with $\delta$ denote fluctuations.
Under these assumptions, we obtain EoMs of energy density fluctuations and flow rapidity fluctuations:
\begin{equation}
\frac{\partial}{\partial \tau}\delta e + \frac{1}{\tau}\left(\delta e + \delta p + \delta \Pi - \delta \pi \right) + \frac{1}{\tau}\frac{\partial}{\partial \eta_{\mathrm{s}}} \delta y \left(e_{0} + p_{0} + \Pi_{0} - \pi_{0}\right) = 0,\label{eq:1st time EoM}
\end{equation}
\begin{equation}
\frac{\partial}{\partial \tau} \delta y\left(e_{0} + p_{0} + \Pi_{0} - \pi_{0}\right) + \frac{2\delta y}{\tau}\left(e_{0} + p_{0} + \Pi_{0} - \pi_{0}\right) + \frac{1}{\tau}\frac{\partial}{\partial \eta_{\mathrm{s}}}(\delta p + \delta \Pi - \delta \pi) = 0.\label{eq:1st space EoM}
\end{equation}
So far, we have not assumed any specific forms of constitutive equations for shear pressure $\pi \equiv \pi^{00} - \pi^{33}$ and bulk pressure $\Pi$ in the course of derivation.
Therefore, we employ the simplest causal constitutive equations \cite{Israel-1, Israel-2} including noise terms $\xi_{\pi}$ and $\xi_{\Pi}$ and linearise these equations following the same prescription as explained above:
\begin{equation}
\delta \pi + \tau_{\pi0} \frac{\partial}{\partial \tau} \delta \pi + \delta \tau_{\pi} \frac{d}{d \tau} \pi_{0} = \frac{4\delta \eta}{3\tau} + \frac{4\eta_{0}}{3\tau}\frac{\partial}{\partial \eta_{\mathrm{s}}} \delta y + \xi_{\pi}\label{1st shear IS final},
\end{equation}
where transport coefficients $\eta$ and $\tau_{\pi}$ are shear viscosity and relaxation time for shear pressure, respectively.
Note that the specific form of the equation of bulk pressure $\Pi$ is quite similar to that of shear pressure $\pi$.
The fluctuations of hydrostatic pressure $\delta p$ and transport coefficients $\delta \eta$ and $\delta \tau_{\pi}$ can be converted from energy density fluctuations once we assume models of EoS and transport coefficients.

We next set the power of noises and its probability distribution.
The noise term $\xi_{\pi}$ obeys the following FDR \cite{Hirano} in the Milne coordinate neglecting possible corrections due to the dynamical evolution of backgrounds \cite{Murase}:
\begin{equation}
\langle \xi_{\pi}(\tau, \eta_{\mathrm{s}}) \rangle = 0, \quad
\langle \xi_{\pi}(\tau, \eta_{\mathrm{s}})\xi_{\pi}(\tau^{\prime}, \eta_{\mathrm{s}}^{\prime}) \rangle = \frac{8\eta_{0} T_{0}}{3\Delta x \Delta y} \delta(\tau-\tau^{\prime})G(\eta_{\mathrm{s}}-\eta_{\mathrm{s}}^{\prime}),
\end{equation}
where the delta functions in transverse plane and in $\eta_{\mathrm{s}}$ direction are replaced with $1/\Delta x \Delta y$ and Gaussian function, $G(\eta_{\mathrm{s}}-\eta_{\mathrm{s}}^{\prime})$, with standard deviation $\sigma_{\eta_{\mathrm{s}}}$, respectively.

We employ two models of the EoS: the conformal EoS, $p=\frac{1}{3}e$, with the degrees of freedom, $d = 47.5$, for massless $N_{f} = 3$ QCD  and  a parametrization of lattice EoS results \cite{lattice}.
In this study we neglect the bulk pressure $\Pi$ for simplicity, which is justified when we use the conformal EoS.
As for the transport coefficients, we employ the specific shear viscosity $\eta/s = 1/4\pi$ \cite{Kovtun} and relaxation time for shear pressure $\tau_{\pi} = (2-\text{ln}2)/2\pi T$ \cite{Baier}.

\section{Results}
\label{sec-2}
Since the equations to be solved are the first order differential equations in time, we need to assign initial conditions for $e_{0}$, $\pi_{0}$, $\delta e$, $\delta \pi$ and $\delta y$.
We start hydrodynamic simulations at initial time $\tau_{\mathrm{ini}} = 1 \;\mathrm{fm}$ with initial conditions, $e_{0} = 10 \;\mathrm{GeV/fm^{3}}$, $\pi_{0} = 4\eta_{0}/3\tau_{\mathrm{ini}} \;\mathrm{GeV/fm^{3}}$, $\delta e = 0 \;\mathrm{GeV/fm^{3}}$, $\delta \pi = 0 \;\mathrm{GeV/fm^{3}}$ and $\delta y = 0$.

\begin{figure}[h]
  \begin{minipage}[b]{0.49\linewidth}
    \centering
    \includegraphics[width=1.1\linewidth,clip]{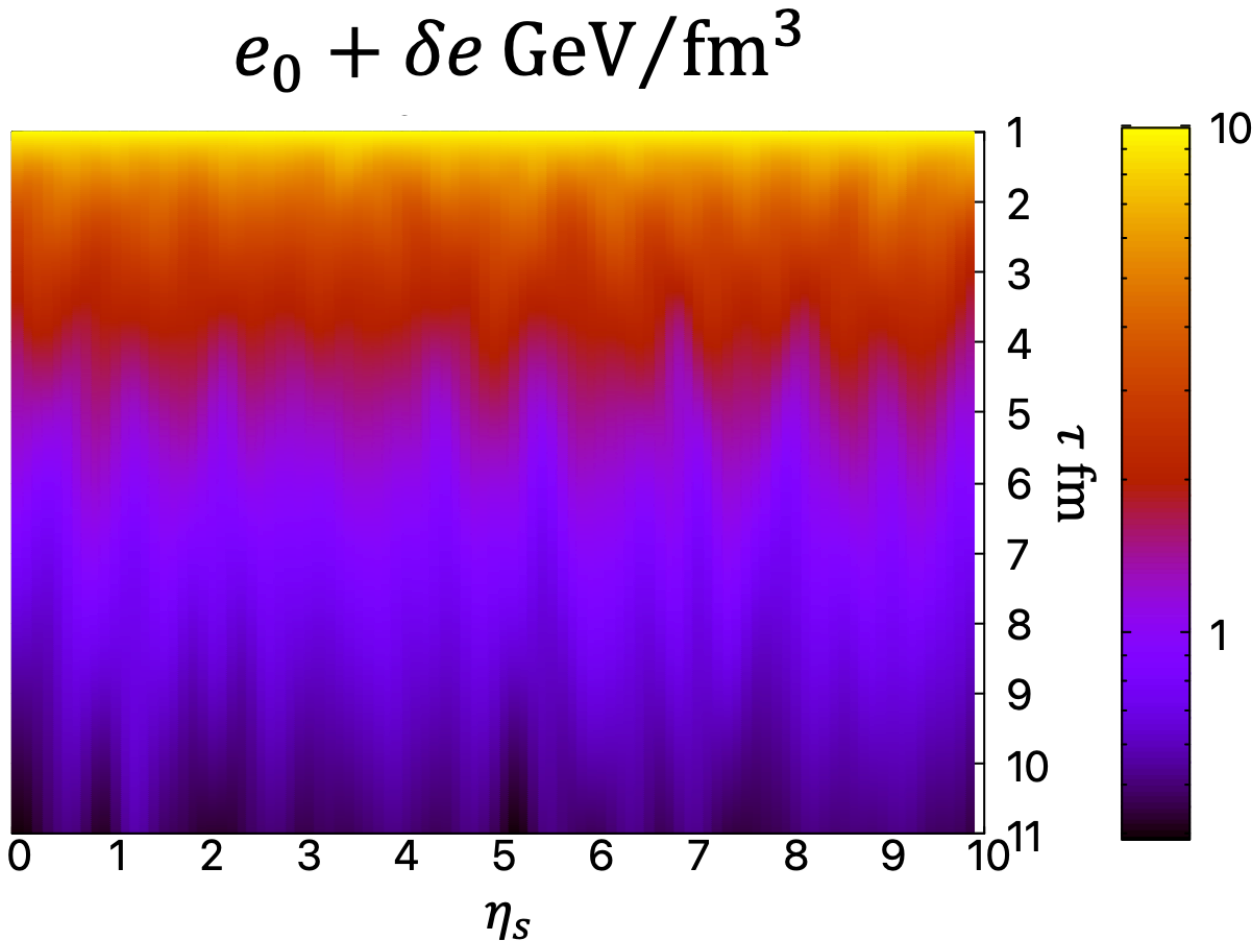}
  \end{minipage}
  \begin{minipage}[b]{0.49\linewidth}
    \centering
    \includegraphics[width=\linewidth,clip]{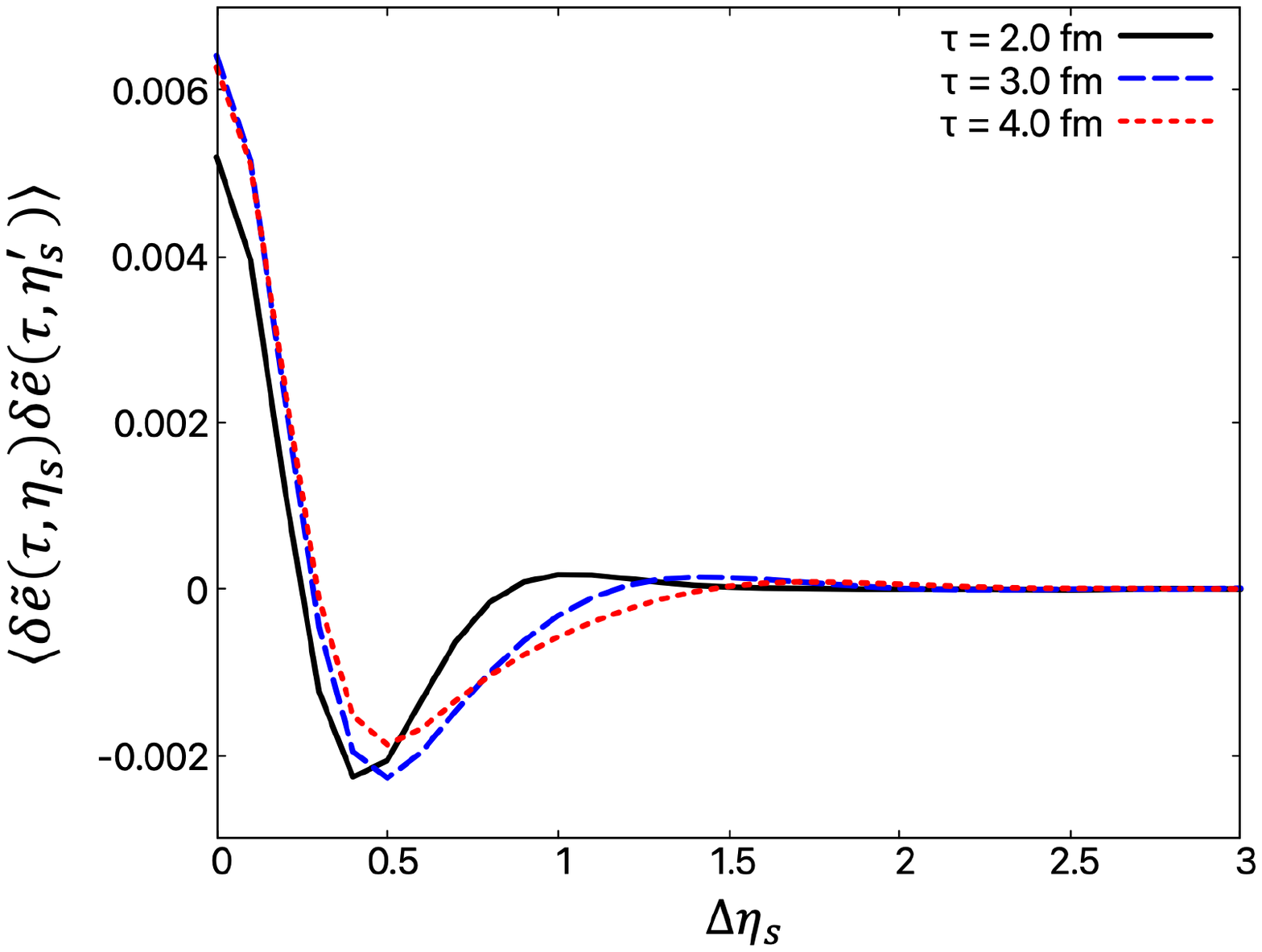}
  \end{minipage}
  \caption{(Left) Space-time evolution of total energy density distribution from one sampled event. (Right) Time evolution of correlation of energy density fluctuations as a function of space-time rapidity gap $\Delta \eta_{\mathrm{s}} \equiv |\eta_{\mathrm{s}}-\eta_{\mathrm{s}}^{\prime}|$ at $\tau =2$, $3$ and $4$ fm.
  The total 10,000 events are averaged to obtain the correlation.
  The conformal EoS is assumed for both results.}
\label{fig:evolution}
\end{figure}
Figure~\ref{fig:evolution} (left) shows  space-time evolution of total energy density distribution from one sampled event.
The background energy density decreases rapidly from the initial value 
due to the rapid expansion of the system.
Remarkably, streak-like structure appears through the time evolution and is kept until the final time $\tau = 11 \;\mathrm{fm}$.
It means that the pattern of the energy density distribution is almost frozen and could carry the information of the early stage.
One possible reason of such a structure formation is that the interplay between the diffusion of fluctuations and the effect of stretching fluctuations due to the rapid expansion of the system.
Figure \ref{fig:evolution} (right) shows time evolution of two-point correlation of normalised energy density fluctuations, $\delta\tilde{e} \equiv \delta e/e_{0}$.
Correlation of normalised energy density fluctuations rapidly grows up around the origin $\Delta \eta_{\mathrm{s}} \sim 0$ and a dip appears at $\Delta \eta_{\mathrm{s}} \sim 0.5$.
This behavior is plausible from the viewpoint of the conservation law. It is clearly seen that the information propagates in $\Delta \eta_{\mathrm{s}}$ direction as time passes.

\begin{figure}[h]
  \begin{minipage}[b]{0.5\linewidth}
    \centering
    \includegraphics[width=\linewidth,clip]{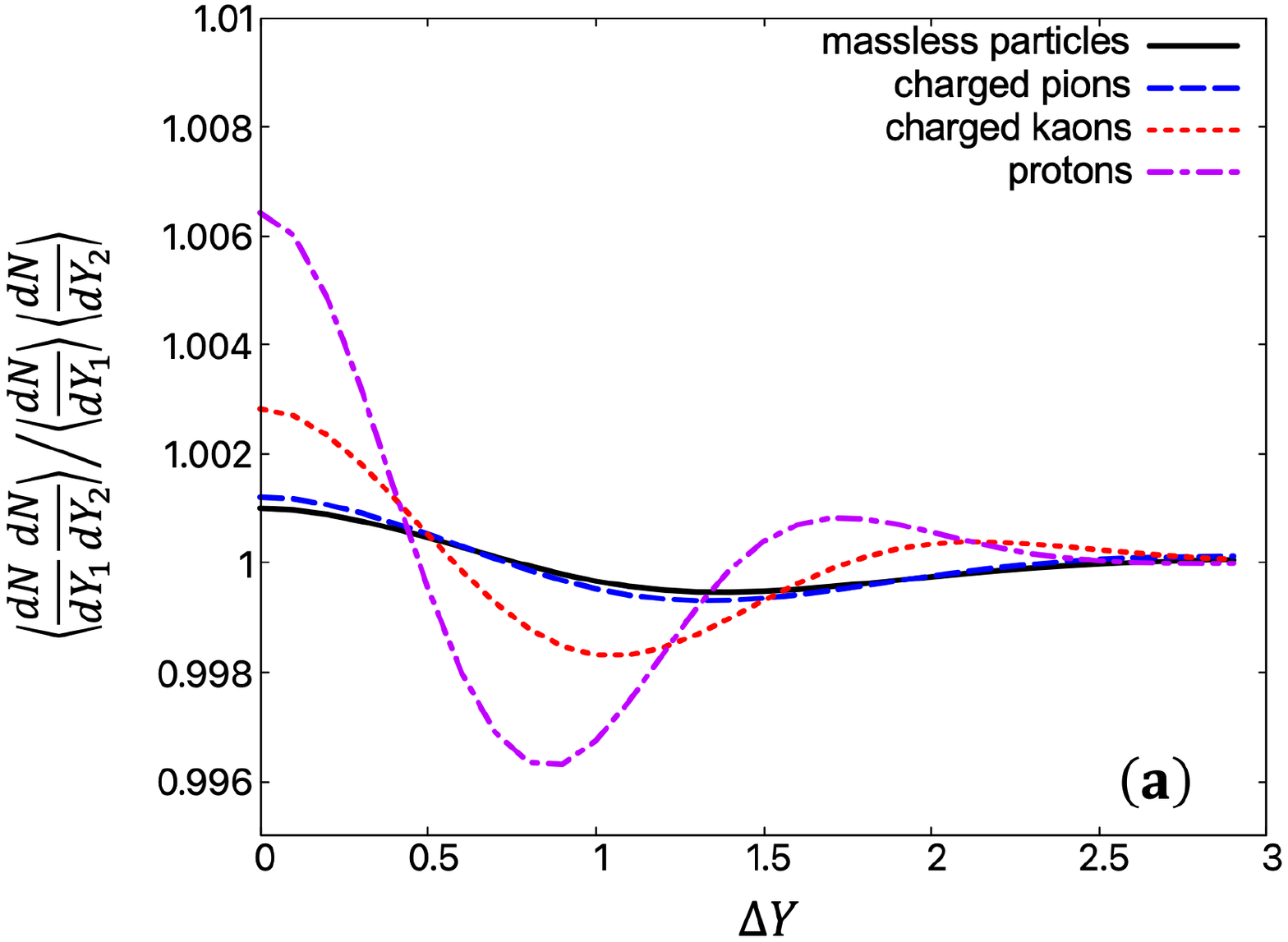}
  \end{minipage}
  \begin{minipage}[b]{0.5\linewidth}
    \centering
    \includegraphics[width=\linewidth,clip]{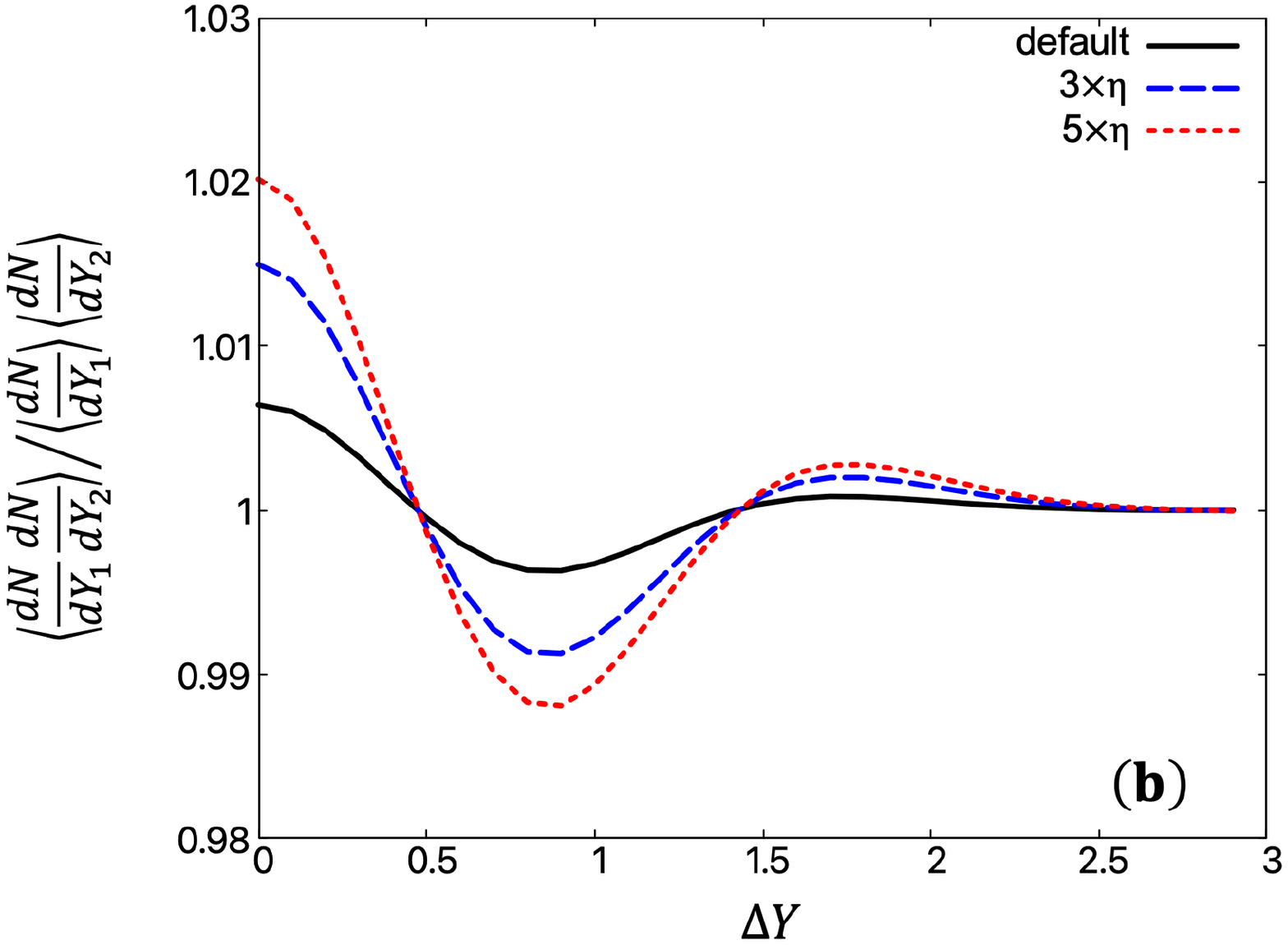}
  \end{minipage}
\caption{Normalised two-particle correlation functions as functions of rapidity gap $\Delta Y \equiv |Y_{1} - Y_{2}|$. 
(a) Hadron species dependence (massless particles,  charged pions, charged kaons and protons).
(b) Shear viscosity dependence of proton two-particle correlation functions (default, 3 times larger $\eta$ and 5 times larger $\eta$).
Isochronous freezeout at $\tau = 10.0 \;\text{fm}$ with lattice EoS \cite{lattice} is performed in both results.
The total 10,000 events are  averaged to obtain the correlation functions.}
\label{fig:particle_correlation}
\end{figure}
To see how correlations of fluctuations of thermodynamic variables and flow rapidity are inherited by two-particle correlations,
we calculate momentum distribution of hadrons via the Cooper-Frye formula \cite {C-F} assuming the Boltzmann distribution with viscous correction \cite{Teaney, Monnai}. 
Figure~\ref{fig:particle_correlation} shows normalised two-particle correlations $\left< \frac{dN}{dY_{1}}\frac{dN}{dY_{2}} \right> / \left< \frac{dN}{dY_{1}} \right> \left< \frac{dN}{dY_{2}} \right>$ as functions of rapidity gap $\Delta Y \equiv |Y_{1}-Y_{2}|$ with various settings, here subscripts 1 and 2 are labels of particle 1 and 2, respectively.
As seen in Fig.~\ref{fig:particle_correlation} (a), the pattern of the correlations is seen more clearly for heavier hadrons. It indicates that the heavier hadrons are better probes of two-particle correlations.
In Fig.~\ref{fig:particle_correlation} (b), one sees viscosity enhances the correlations. 
Note that relaxation time turns out to suppress the correlations (not shown).
The magnitude of two-particle correlations is highly sensitive to the transport coefficients. Nevertheless the positions of dips and bumps do not depend on them.
We also find that  the shape of two-particle correlations is different from each other according to the choice of the EoS.
It indicates that two-particle correlations include  information of not only transport coefficients but also the EoS.

\section{Summary}
\label{sec-3}
We developed a framework of causal hydrodynamic fluctuations in one-dimensional expanding system by means of linearisation.
Through the description of space-time evolution of thermodynamic variables, we found a frozen structure of thermodynamic variables.
From the analysis of the structure, it could be possible to extract the information of the early stage of hydrodynamic evolution.
We found that two-particle correlations as functions of rapidity gap are more enhanced for heavier hadrons and that the magnitude of two-particle correlations is sensitive to the properties of the medium such as viscosity, relaxation time and EoS.
These results provide us with an opportunity for multidimensional analysis of the QGP properties created in the relativistic heavy ion collisions.

\section*{Acknowledgement}
\label{sec-4}
The work by T.H. was partly supported by JSPS KAKENHI Grant No.~JP19K21881.

%
%
%

\end{document}